\documentclass[a4paper]{article}

\usepackage{INTERSPEECH2020}
\usepackage{multirow}
\usepackage{hyperref}

\def\F{{\mathbf{F}}}
\def\Q{{\mathbf{Q}}}
\def\K{{\mathbf{K}}}
\def\V{{\mathbf{V}}}

\title{Sams-Net: A Sliced Attention-based Neural Network \\ for Music Source Separation}
\name{Tingle Li$^{1,3}$, Jiawei Chen$^3$, Haowen Hou$^4$, Ming Li$^{1,2}$}
\address{
  $^1$Data Science Research Center, Duke Kunshan University, Kunshan, China \\
  $^2$School of Computer Science, Wuhan University, Wuhan, China \\
  $^3$School of Computer Science and Technology, Tiangong University, Tianjin, China\\
  $^4$AI Platform Department, Tencent Inc., Shenzhen, China}
\email{ming.li369@dukekunshan.edu.cn}

\begin{document}

\maketitle
\begin{abstract}
Convolutional Neural Network (CNN) or Long short-term memory (LSTM) based models with the input of spectrogram or waveforms are commonly used for deep learning based audio source separation. In this paper, we propose a Sliced Attention-based neural network (Sams-Net) in the spectrogram domain for the music source separation task. It enables spectral feature interactions with multi-head attention mechanism, achieves easier parallel computing and has a larger receptive field compared with LSTMs and CNNs respectively. Experimental results on the MUSDB18 dataset show that the proposed method, with fewer parameters, outperforms most of the state-of-the-art DNN-based methods.
\end{abstract}
\noindent\textbf{Index Terms}: Sliced Attention, Music Source Separation, Music Information Retrieval

\section{Introduction}

The ``cocktail party effect" was first proposed by Cherry \cite{cherry1953some}: How does the human brain separate a conversation from the surrounding noise. Later, Bregman \cite{bregman1994auditory} tried to study how the human brain analyzed the complex auditory signal, and proposed a framework for it. By the early 21st century, Roman \cite{roman2006pitch} attempted to simulate the brain's ability of source separation by means of algorithms, which served as the main framework of source separation right now. When it comes to music source separation, the first unsupervised method is \cite{li2007separation}. Recently, the supervised methods, especially the deep learning-based methods \cite{huang2014singing, jansson2017singing}, have achieved promising performance for this task.

Music is produced by assembling sound from multiple individual instruments called stems. The goal of music source separation is to recover those individual stems from the mixed signal \cite{vincent2005musical}. In the SiSEC 2018 campaign \cite{stoter20182018}, those individual stems were grouped into four categories: vocals, drums, bass and other. Given a song which is a mixture of these four sources, our goal is to separate it into four parts that correspond to the original sources. Unlike the speech separation task in which each single source is independent to each other, there are many similar music pieces recurring in the same source and among different sources in a song. This characteristic brings a great challenge for music source separation.

Most of the music source separation models can be categorized into two main categories, namely the spectrogram based methods \cite{huang2014singing, jansson2017singing, stoter2019open}, and the waveform based methods \cite{oord2016wavenet, stoller2018wave, Demucs2019}. DNN based music source separation models are mainly based on three network architectures: Fully Connected Network (FCN) \cite{simpson2015deep}, Convolutional Neural Networks (CNN) \cite{jansson2017singing, stoller2018wave} and Long Short-Term Memory (LSTM) \cite{huang2014singing}. Recently, CNN and LSTM have been combined to achieve state-of-the-art performance for music source separation \cite{Demucs2019, takahashi2018mmdenselstm}. However, both CNN and LSTM have certain limitations. For CNN, the problem lies in the requirement of a large receptive field for source separation task \cite{luo2016understanding}. Although deeper CNNs are able to obtain a larger receptive field, the increase of parameters makes training difficult. Another way to expand the receptive field is to apply pooling layers and aggregate the context, which, however, result in the loss of spectral detail. To solve this problem, CNN with a multi-scale structure is proposed \cite{stoller2018wave}. This multi-scale structure adopts downsampling (i.e., max-pooling layer) to get low resolution feature maps from which the upsampling layers are employed to recover the original resolution. For LSTM \cite{stoter2019open}, the time-dependent property prevents the parallel computation and makes inference time-consuming. Furthermore, the long-term dependency issue is not well addressed in LSTMs according to the previous study in \cite{hochreiter2001gradient}.

The attention mechanism \cite{bahdanau2014neural} is a recent advance in neural network modeling. It automatically learns feature interactions from data without any human domain knowledge. Recently, Vaswani et al. proposed a novel neural network structure - Transformer \cite{vaswani2017attention}, which uses only the attention mechanism structure to obtain the state-of-the-art result in the English-French translation task. Using the attention mechanism, the Transformer is a structure that can automatically capture sequence distribution. It allows the network to learn which part of the input sequence is important, and which ones are not. Experimental results show that the Transformer is more suitable for processing sequences than LSTMs, because it can solve long-term dependency problems better than LSTMs \cite{vaswani2017attention}. Also, since the attention mechanism has no time-dependent limitation for calculating, the Transformer can be computed in parallel easily. Furthermore, the Transformer has a larger receptive field compared with CNNs with the same number of layers. Shortly after that, a well-known framework called BERT \cite{devlin2018bert} was proposed to reduce the gap between the pre-training word embedding and the downstream specific Natural Language Processing (NLP) task.

Motivated by the successful application of attention mechanism, we try to study the capability of attention mechanism in the source separation task, and propose a neural network for music source separation named Sams-Net, which applies the proposed Sliced Attention mechanism to the spectrogram domain. Besides, although many time-domain models have achieved better signal-to-distortion ratio (SDR) \cite{vincent2006performance} than those of the spectrogram domain, modeling in the time domain does not produce good quality speech \cite{pandey2018new}. So we are interested in building our model in the spectrogram domain like \cite{stoter2019open}, which only feeds the magnitude spectrogram into the model. Experimental results show that our model has achieved a new state-of-the-art result.

The rest of the paper is organized as follows. Section 2 introduces the problem formulation of the music source separation task. Section 3 illustrates the core modules of the proposed Sams-Net. Experimental setup and results are reported in Section 4, while conclusions are drawn in Section 5.

\section{Music Source Separation}
Music source separation is to extract source signals from the musical mixture. Specifically, a stereo mixture $\mathbf{x} \in \mathbb{R}^{2 \times \widetilde{T}}$ can be expressed as a linear combination of $c$ source signals $\mathbf{s} \in \mathbb{R}^{2 \times \widetilde{T}}$:
\begin{equation}
\mathbf{x}(t) = \sum_{i=1}^c \mathbf{s}_i(t)
\end{equation}

To perform music source separation in spectrogram domain, the Short-time Fourier Transform (STFT) $\mathbf{S} \in \mathbb{R}^{2 \times T \times F}$ for each source signal $\mathbf{s}$ is calculated to get the mixed spectrogram $\mathbf{X} \in \mathbb{R}^{2 \times T \times F}$ during training:
\begin{equation}
\mathbf{X}(t, f) = \sum_{i=1}^c \mathbf{S}_i(t,f)
\end{equation}
where $T$ and $F$ are the dimension of the time and frequency axis respectively.

The separation network then takes the magnitude spectrogram $|\mathbf{X}|$ as input to estimate a mask $\mathbf{M} \in \mathbb{R}^{2 \times T \times F}$ for each source. The reconstruction of time domain source $\mathbf{\widehat{s}}$ is accomplished by calculating the inverse STFT (ISTFT) for the estimated spectrogram, which is:
\begin{equation}
\mathbf{\widehat{s}}_{i}(t)=\mathrm{ISTFT}(|\mathbf{X}(t, f)| \odot \mathbf{M}_{i}(t, f)) \times e^{\angle \mathbf{X}(t, f)j})
\end{equation}
where $\angle \mathbf{X}(t, f)$ is the phase of mixed musical segment, and $\odot$ is element-wise multiplication.

To this end, the learning objective is to minimize the audio based squared ${l_2}$-norm \cite{stoller2018wave} between the original and estimated sources:
\begin{equation}
\mathcal{L}=\underset{\mathbf{M}}{\arg \min } \sum_{i=1}^c\left\|\mathbf{s}_{i}(t) - \mathbf{\widehat{s}}_{i}(t) \right\|_{2}^{2}
\end{equation}

\begin{figure}[h]
\centering
\includegraphics[width=\linewidth]{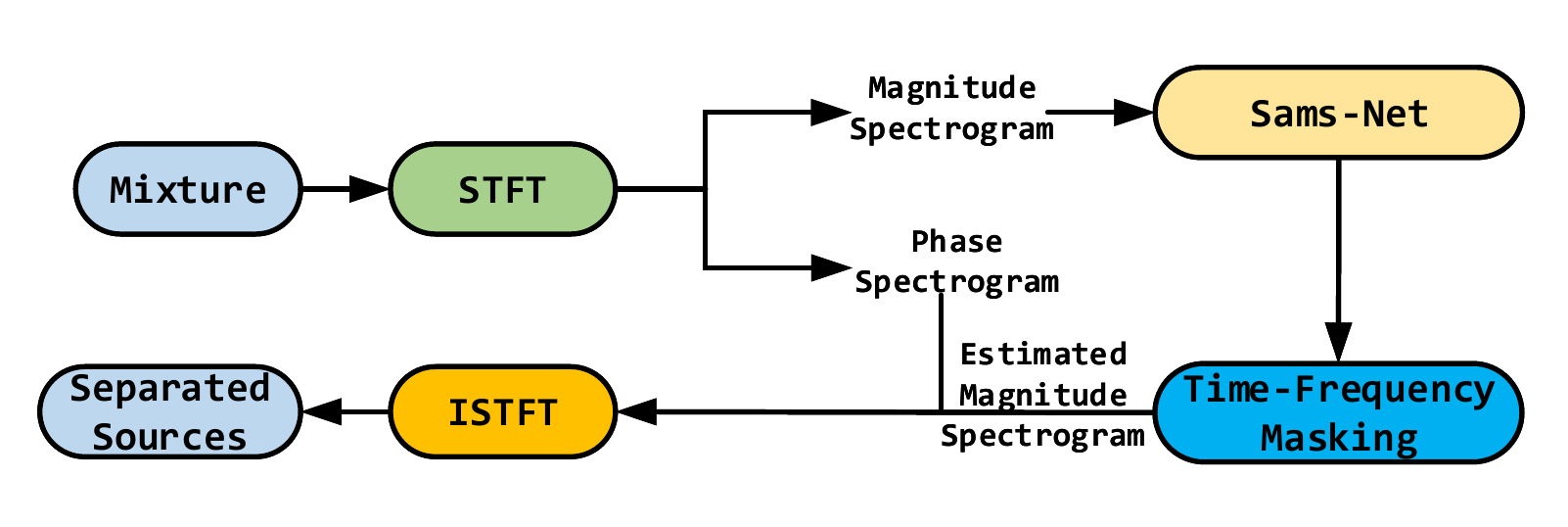}
\caption{The flow chart of our separation system.}
\label{fig:source_separation}
\end{figure}

Figure \ref{fig:source_separation} shows the flowchart of a typical music source separation system on the spectrogram domain.

\section{Model Description}

\begin{figure}[tb]
  \centering
  \includegraphics[scale=0.43]{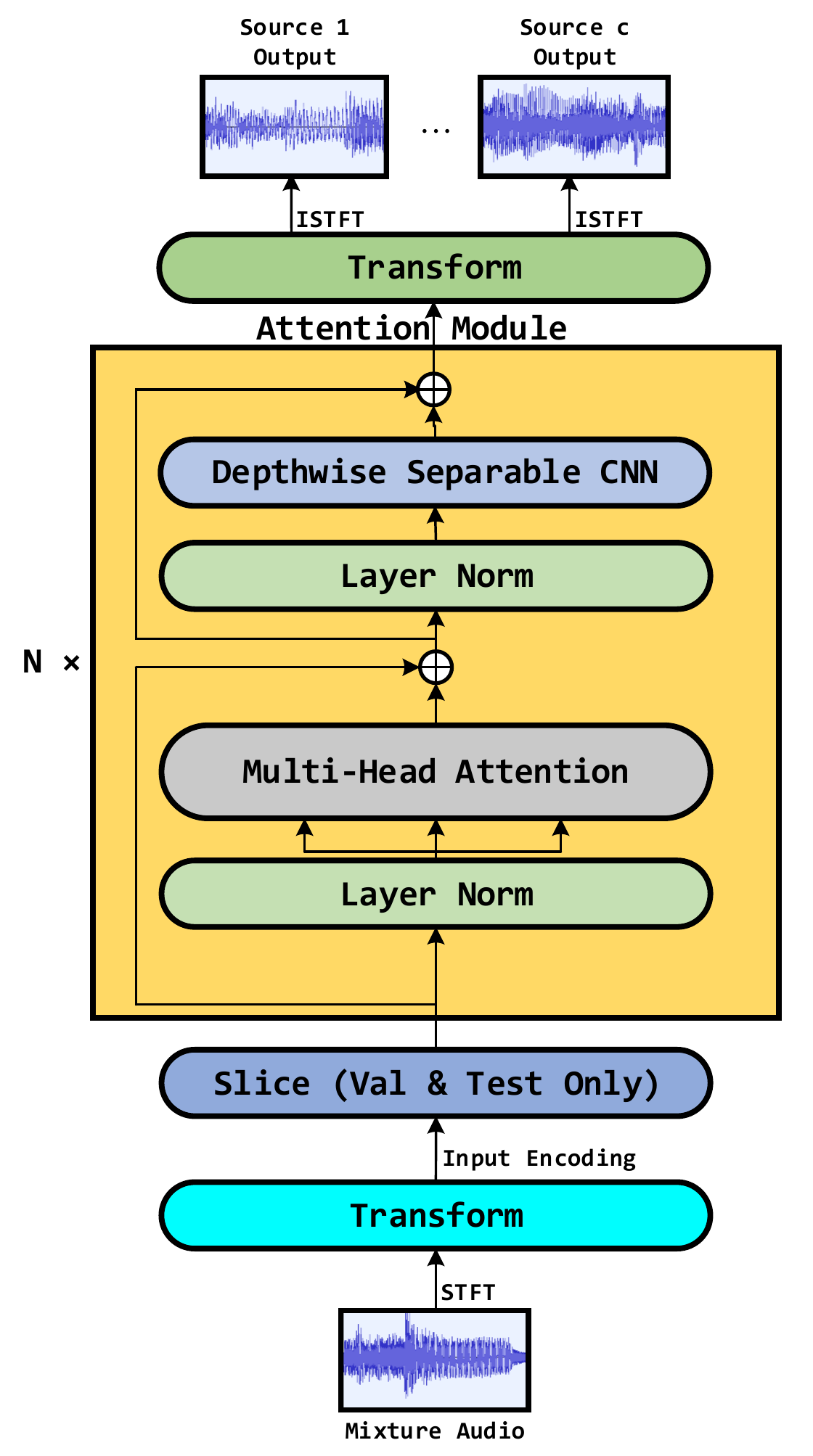}
  \caption{Model architecture of Sams-Net.}
  \label{fig:sams_net}
\end{figure}

This section describes our proposed Sams-Net for music source separation, as shown in Figure \ref{fig:sams_net}. It includes two transform modules, a slice module and $N$ attention modules. The first transform module is a CNN layer that takes the magnitude spectrogram as input and expands the channel of the feature map to $C$, while the second one is a transpose CNN layer that aggregates the channel dimension of the feature map from $C$ to $2$ and generates the estimated masks. The slice module (section 3.3) is only applicable to the validation and test stage, not the training stage. The attention block consists of a multi-head scaled dot-product attention layer (section 3.1-3.3), a depth-wise separable CNN layer (section 3.4) and several layer normalization layers (section 3.5).

\subsection{Scaled Dot-Product Attention}

The common attention methods include dot-product \cite{vaswani2017attention}, concatenation \cite{anderson2018bottom}, perceptron \cite{oktay2018attention}, etc. In this study, we apply scaled dot-product attention to each channel of a feature map. Specifically, for a given feature map in feature space of $\mathbb{R}^{C \times T \times F}$, three CNN layers with kernel size of $1 \times 1$ are employed separately to transform this feature map into queries $\Q$, keys $\K$, and values $\V \in \mathbb{R}^{C \times T \times F}$. Then, for each feature channel $c$, dot products of the $t^\mathrm{th}$ time frame of query $\Q_{c,t} \in \mathbb{R}^{F}$ and all keys $\K_c \in \mathbb{R}^{T\times F}$ are calculated as attention weights. Softmax is applied to the attention weights before aggregating the values $\V_{c}$. In practice, we compute the attention function on a set of queries $\Q_{c}$ simultaneously:
\begin{equation}
\mathrm{Attention}({\Q_c},{\K_c},{\V_c}) = \mathrm{softmax}\left( {\frac{{{\Q_c}{{\K_c}^T}}}{{\sqrt{C}}}} \right){\V_c}
\end{equation}
where $ 1 / \sqrt{C} $ is a regulating term to avoid large inner product as well as gradient vanishing. We denote the scaled dot-product attention for the whole feature map as $\mathrm{Attention}({\Q},{\K},{\V})$.

\subsection{Multi-Head Attention}
To further exploit the modeling ability of attention mechanism, we apply multi-head attention \cite{vaswani2017attention} on top of the scaled dot-product attention. In this study, the output feature map $\F$ of the transform block is firstly fed into three convolutional layers to produce queries $\Q_h$, keys $\K_h$, and values $\V_h \in \mathbb{R}^{C \times T \times F}$ for the $h^\mathrm{th}$ head. Then, the scaled dot-product attention is applied for each of the $H$ heads, and the outputs are concatenated along the channel axis:
\begin{equation}
\begin{aligned}
& \mathrm{MultiHead}({\Q},{\K},{\V}) = \mathrm{Concat}(\mathrm{head}_1, \cdots, \mathrm{head}_H), \\
& \mathrm{head}_h = \mathrm{Attention}(\mathrm{Conv}_h^Q(\F),\mathrm{Conv}_h^K(\F),\mathrm{Conv}_h^V(\F))
\end{aligned}
\end{equation}
where $\mathrm{Conv}_h^Q, \mathrm{Conv}_h^K, \mathrm{Conv}_h^V$ are CNN layers to produce queries, keys, and values for $h^\mathrm{th}$ head.

Finally, an additional CNN layer with the kernel size $3 \times 3$ is employed to recover the channel dimension of the multi-head attentive output from $H \times C$ to $C$.

\subsection{Sliced Attention}
Due to the typically long duration of a song in the practical application, we apply slice operation to the magnitude spectrogram before attention operation.

Specifically, the magnitude spectrogram is sliced into $I$ chunks without overlap, yielding $T/I$ frames per chunks. Multi-head attention is applied to these $I$ chunks separately. The resulting of $I$ attention values are concatenated along the time axis as the final output SA:
\begin{equation}
\begin{aligned}
\mathrm{SA} & = \mathrm{Concat}(\mathrm{slice}_1, \mathrm{slice}_2, \cdots, \mathrm{slice}_I) \\
\mathrm{slice}_i & = \mathrm{MultiHead}(\Q_{i}, \K_{i}, \V_{i})
\end{aligned}
\end{equation}
where $\Q_{i}, \K_{i}$ and $\V_{i}$ are queries, keys and values for chunk $i$.

\begin{figure}[tb]
  \centering
  \includegraphics[scale=0.45]{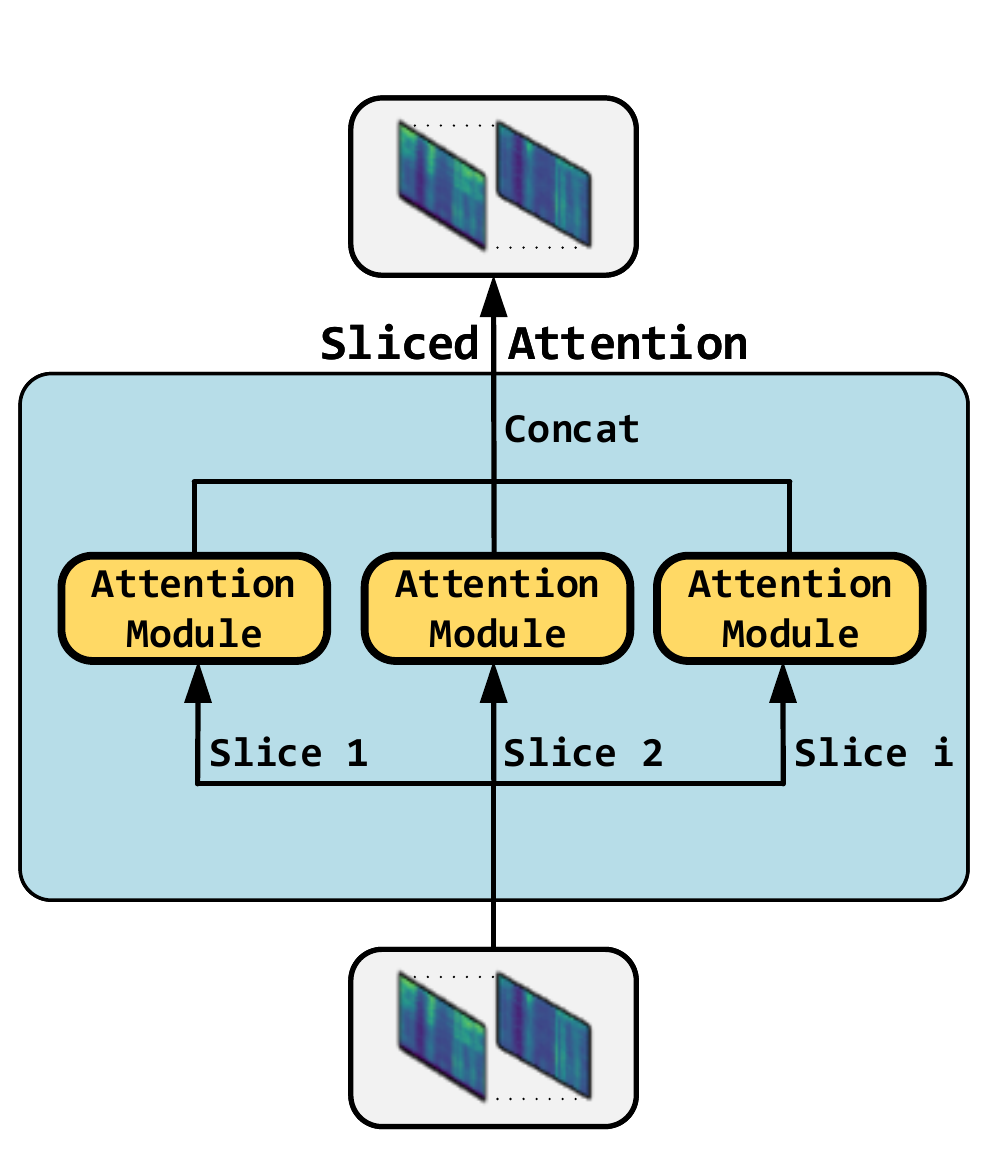}
  \caption{Model architecture of our Sliced Attention module.}
  \label{fig:sliced_attention}
\end{figure}

With the sliced operation, the scope of attention is narrowed down to the intra-chunk features. We define the slice operation with the multi-head scaled dot-product attention as sliced attention. Figure \ref{fig:sliced_attention} shows a flowchart of the sliced attention. The feature dimension of the output remains unchanged after sliced attention.

The reason we apply a slice operation lies in the inherent data pattern of songs. There are recurring elements \cite{paulus2010state}, such as notes, pitch, timbre, and chords within a song. Also, musical style or pattern always changes. For example, the principal musical instrument can suddenly change from drums to bass, followed by a pure human voice. In this case, the sliced attention provides a mechanism for the network to focus a small piece of the song with the same musical style, without taking other parts which are irrelevant to the current one.

The slice operation is applied to a whole song at validation and test stages. During training, a short chunk is randomly sliced within a song at each training step.

\subsection{Depthwise Separable CNN}
Considering the large size of our model, we choose lightweight depthwise separable CNN \cite{chollet2017xception} rather than conventional CNN. Specifically, it has two operations: depth-wise convolution and point-wise convolution.

For depth-wise convolution, each channel of the feature map is applied to a convolutional layer with the kernel size of $3 \times 3$ independently. Different from the conventional CNN, one convolutional kernel of the depth-wise convolutional layer is responsible for one channel. For point-wise convolution, the output of the depth-wise convolution operation is fed to a convolutional layer with the kernel size of $1 \times 1$, which aims to further aggregate the information among different channels of the feature map.

\subsection{Layer Norm}
In the proposed Sams-Net, layer normalization \cite{ba2016layer} is performed after the sublayers, i.e., the sliced attention layers, and the depthwise separable CNN layers.

As shown in Figure \ref{fig:sams_net}, residual connections are also applied between two normalization layers, i.e., shortcut connection is inserted between the inputs of the previous and current normalization layer. The computing formular is as follows:
\begin{equation}
x + \mathrm{Sublayer}(\mathrm{LayerNorm}(x))
\end{equation}
where $\mathrm{Sublayer}(x)$ is a function implemented by the sublayer itself.

\section{Experimental Results}

\subsection{Experimental setup}
We evaluated Sams-Net on MUSDB18 dataset \cite{rafii2017musdb18}, which is prepared for the SiSEC 2018 campaign \cite{stoter20182018}. MUSDB18 has 100 and 50 songs in the training and test set, respectively. In this dataset, each song is a mixture with four sources, i.e., vocals, bass, drums, and other, all of which are recorded in stereo format with the sampling rate of 44.1 kHz. During the training stage, we randomly select 6 seconds from each training track as the training data. In the validation stage, we use the recommended tracklist provided by \cite{rafii2017musdb18} as the validation data.

The STFT used in our system is computed based on a 93 ms Hamming window with 75\% overlap between frames and a 4096-point discrete Fourier transform. Also, data augmentation \cite{uhlich2017improving} is used during training. We apply random gains between 0.25 and 1.25 to all sources before mixing. Besides, the channels of the stereo mixtures are randomly swapped. For the evaluation on MUSDB18, we used the museval package \cite{stoter20182018} and BSSEval v4 toolbox \cite{vincent2006performance} for a fair comparison with previously reported results. The average of the median signal-to-distortion-ratio (SDR) of each song in the test set is used as the evaluation metric.

\begin{table*}[!ht]
\setlength{\tabcolsep}{3.5mm}{
\begin{center}
\renewcommand{\arraystretch}{1.1}
\caption{A comparison SDR metric of our proposed method with other models on the test set of MUSDB18 dataset.}
\begin{tabular}{llllllll}
\toprule
\multirow{2}{*}{\textbf{Model} } & \multirow{2}{*}{\textbf{Domain} } & \multirow{2}{*}{\textbf{\# Param}} & \multicolumn{5}{l}{~ ~ ~ ~ ~ ~ ~ ~ ~ ~ ~ ~ ~ ~ ~ ~\textbf{Test SDR (dB)} }                           \\
\cline{4-8}
            &                     &                                   & Vocals         & Drums          & Bass           & Other          & Average              \\
\midrule
IRM oracle                       & N/A                & N/A               & 9.43           & 8.45           & 7.12           & 7.85           & 8.21             \\
\midrule
DeepConvSep~\cite{chandna2017monoaural}                      & Spectrogram        & 0.32M               & 2.37           & 3.14           & 0.17           & -2.13          & 0.89             \\
WaveNet~\cite{lluis2018end}                          & Waveform              & 3.30M            & 3.35           & 4.13           & 2.49           & 2.60            & 2.60              \\
Wave-U-Net~\cite{stoller2018wave}                       & Waveform             & 10.20M             & 3.25           & 4.22           & 3.21           & 2.25           & 3.23             \\
Spect U-Net~\cite{liu2018denoising}                      & Spectrogram          & 9.84M             & 5.74           & 4.66           & 3.67           & 3.40            & 4.37             \\
Open-Unmix~\cite{stoter2019open}                       & Spectrogram            & 8.90M           & 6.32           & 5.73           & 5.23           & 4.02           & 5.36             \\
Demucs~\cite{Demucs2019}                           & Waveform          & 648.00M                & 6.29           & 6.08           & \textbf{5.83}  & 4.12           & 5.58             \\
Meta-TasNet~\cite{samuel2020meta}                      & Waveform         & 45.50M                 & 6.40            & 5.91           & 5.58           & \textbf{4.19}  & 5.52             \\
MMDenseLSTM~\cite{takahashi2018mmdenselstm}                      & Spectrogram       & 4.88M                & 6.60           & 6.41           & 5.16           & 4.15 & 5.58~ ~ ~ ~ ~ ~  \\
\midrule
Sams-Net (w/o Attention)               & Spectrogram          & 3.64M             & 4.80 & 4.71 & 3.89           & 3.22           & 4.16    \\
\midrule
\textbf{Sams-Net}                & Spectrogram          & 3.70M             & \textbf{6.61} & \textbf{6.63} & 5.25           & 4.09           & \textbf{5.65}    \\
\bottomrule
\end{tabular}
\end{center}}
\end{table*}

We train the model based on PyTorch framework \cite{paszke2019pytorch} with 2 NVIDIA TITAN RTX GPU. The parameters of the network are updated using the Adam optimizer \cite{kingma2014adam} with an initial learning rate of 0.0001. We terminate the training process when the validation loss no longer descends after 140 epochs. Due to the limitation of the GPU memory, we can only train 3 Sliced Attention modules with 2 heads and 64 channels of the CNN feature maps.

\subsection{Results and Discussions}
Table 2 shows the SDR metrics with different numbers of slices used in Sams-Net. The observation is that models with the proposed slice operation ($I$ = 2,4,8,12) outperform the model without slice operation ($I$ =1). As the number of slices increases from 1 to 12, the four sources' SDR metrics also increase. However, performance degrades when the slice number increases above 12. The reason may be that each chunk's duration is too short to contain sufficient key features for attentive value learning with a large number of slices. For the MUSDB18 dataset, the optimal slice number is around 12, but this does not apply to other datasets, especially for those with unknown sources.

\begin{table}[tb]
\centering
\caption{A comparison SDR metric of Sams-Net among several slice numbers (in dB), .}
\begin{tabular}{l|lllll}
\toprule
\textbf{Slices} & \textbf{Vocals} & \textbf{Drums} & \textbf{Bass} & \textbf{Other} & \textbf{Average}  \\
\midrule
1               & 6.42            & 6.55           & 5.21 & 3.97           & 5.53              \\
2               & 6.44            & 6.59           & \textbf{5.32} & 4.04           & 5.60              \\
4               & 6.42            & 6.57           & 5.29 & 4.08           & 5.59              \\
8               & 6.61            & 6.56           & 5.23          & 4.06           & 5.62              \\
\textbf{12}     & \textbf{6.61}   & \textbf{6.63}  & 5.25          & \textbf{4.09}  & \textbf{5.65}     \\
16              & 6.51            & 6.45           & 5.19          & 4.01           & 5.54              \\
18              & 6.57            & 6.37           & 4.94          & 3.88           & 5.44              \\
24              & 6.26            & 5.98           & 4.55          & 3.68           & 5.12              \\
\bottomrule
\end{tabular}
\end{table}

As shown in Table 1, the performance of Sams-Net without the attention module decreases significantly, which means the effectiveness of the attention mechanism used here. Also, we compare our model against previously published and state-of-the-art models for the MUSDB18 dataset (models trained with extra data are not listed here). The SDR results are either taken from the SiSEC 2018 \cite{stoter20182018} evaluation scores \cite{stoller2018wave, takahashi2018mmdenselstm, liu2018denoising}  or from the related papers \cite{stoter2019open, Demucs2019, chandna2017monoaural, lluis2018end, samuel2020meta}. The Ideal Ratio Mask oracle (IRM oracle) \cite{stoter2019open} at the top line computes the best possible mask using the ground truth sources. The proposed Sams-Net outperforms both waveform-based and spectrogram-based models on vocals and drums categories, and achieve the state-of-the-art result on the averaged score. Besides, compared with the best models of Demucs \cite{Demucs2019} and Meta-TasNet \cite{samuel2020meta} on bass and other categories, the proposed network contains 99.4\% and 91.8\% fewer parameters with only 9.9\% and 2.4\% relative performance degradation respectively. We also provide some samples inferred by Sams-Net and other models \cite{stoter2019open, Demucs2019} online\footnote{\url{https://tinglok.netlify.com/files/samsnet/}} for intuitive comparison.

The number of parameters in Sams-Net is about 3.7M, which is smaller than that of most other methods. Compared with the baseline Open-Unmix model \cite{stoter2019open} containing about 8.9M parameters, our model achieves 5.4\% improvement in terms of average SDR. We believe that the increase of parameters of Sams-Net can further enhance the performance.

\section{Conclusions}
In this paper, we propose a sliced attention-based neural network for music source separation, named Sams-Net. It improves the performance of music source separation by discriminating the importance of different feature interactions. We also propose a sliced attention mechanism to learn the importance of each feature interaction from every segment of the magnitude spectrogram. As shown by the experimental results, the proposed Sams-Net, with lesser parameters, achieves better performance than other methods in terms of SDR.

For future works, we intend to model the data in the time-domain to incorporate the phase information. Moreover, as the points of music style change are not uniformly distributed, we can't just simply slice each song equally. We hope to automatically identify the change point of musical style within a song, so that the sliced attention can be performed within a chunk with the same style. Classifier at frame-level may be a possible solution to find the pattern of musical styles for each frame, so that a change point can be determined.

\section{Acknowledgements}
This research is funded in part by the National Natural Science Foundation of China (61773413), Key Research and Development Program of Jiangsu Province (BE2019054), Six talent peaks project in Jiangsu Province (JY-074), Science and Technology Program of Guangzhou, China (202007030011, 201903010040).

\bibliographystyle{IEEEtran}

\bibliography{mybib}

\end{document}